\documentclass[english,superscriptaddress,prl,twocolumn]{revtex4}
\usepackage[toc,page,title,titletoc,header]{appendix}
\usepackage[T1]{fontenc}
\usepackage{enumerate}
\usepackage{xcolor}
\usepackage{graphicx}
\usepackage[latin1]{inputenc}
\usepackage{epstopdf}
\usepackage{amsthm}

\usepackage{amsmath}
\usepackage{shapepar}
\usepackage{color}

\usepackage{amssymb, amsbsy, amsthm}
\makeatletter
\newcommand{\tr}{\mbox{\rm Tr}}

\newtheorem{Theorem}{Theorem}

\usepackage{babel}
\makeatother
\begin{document}
\title{Bounding the amount of entanglement from witness operators}
\author{Liang-Liang Sun}
\email{sun18@ustc.edu.cn}
\affiliation{Department of Modern Physics and National Laboratory for Physical Sciences at Microscale, University of Science and Technology of China, Hefei, Anhui 230026, China}

\author{Xiang Zhou}
\affiliation{Department of Modern Physics and National Laboratory for Physical Sciences at Microscale, University of Science and Technology of China, Hefei, Anhui 230026, China}

\author{Armin Tavakoli}\email{armin.tavakoli@teorfys.lu.se}
\affiliation{Physics Department, Lund University, Box 118, 22100 Lund, Sweden}

\author{Zhen-Peng Xu}\email{zhen-peng.xu@uni-siegen.de}
\affiliation{School of Physics and Optoelectronics Engineering, Anhui University, Hefei 230601, People's Republic of China}

\author{Sixia Yu }\email{yusixia@ustc.edu.cn}
\affiliation{Department of Modern Physics and National Laboratory for Physical Sciences at Microscale, University of Science and Technology of China, Hefei, Anhui 230026, China}

\date{\today{}}
\begin{abstract}
We present an approach to estimate the operational distinguishability between an entangled state and any separable state directly from measuring an entanglement witness. We show that this estimation also implies bounds on a variety of other well-known entanglement quantifiers. This approach for entanglement estimation is then extended to to both the measurement-device-independent scenario and the fully device-independent scenario, where we obtain non-trivial but sub-optimal bounds. The procedure requires no numerical optimization and is easy to compute. It offers ways for experimenters to not only detect, but also quantify, entanglement from the standard entanglement witness procedure.
\end{abstract}


\maketitle

\section{I. Introduction}
Entanglement~\cite{RevModPhys.81.865} is a fundamental  quantum  resource promising  advantages over the classical resources in many information tasks ranging from quantum computation and communication~\cite{Briegel2009} to metrology~\cite{PhysRevLett.67.661, Giovannetti2004}. A fundamental problem is to decide if a given state is entangled or not. If  affirmative, a natural problem is then to quantify the amount of entanglement present in the state. However, both these problems are typically NP-hard~\cite{np2, np1, Huang_2014}, and that has led to the development of a variety of partial entanglement criteria \cite{Horodecki2009, nrev} and computational methods \cite{SDPreview}.

Practical entanglement detection must be reasonably efficient, i.e.~use far fewer local measurements than those needed for state tomography\, while still maintaining an adequate detection power~\cite{Guhne2005, Morelli2023, Liu2022}. The leading tool for such purposes is  entanglement witnesses (EWs)~\cite{Terhal2000}, which are observable quantities that are by convention normalised so that they are positive for all separable states and negative for some entangled states. Every entangled state can in principle be detected by some suitably chosen EW~\cite{RevModPhys.81.865}, but due to its computational hardness\cite{np1, PhysRevLett.88.187904, PhysRevA.69.022308, PhysRevLett.117.210504, PhysRevLett.118.110502}, the common approach is to instead target simple EWs that give only sufficient conditions for entanglement~\cite{nrev}.

Assuming that a device precisely measures the observables prescribed in an EW is usually insufficient. This can for instance be ascribed either to the realistic influence of imperfect control \cite{PhysRevA.86.062325, Morelli2022} or to the presence of uncharacterised devices. A well-known scenario to overcome such issues is the measurement-device-independent (MDI) scenario \cite{PhysRevLett.108.200401, PhysRevLett.110.060405}. EWs designed for the MDI scenario work with uncharacterised  measurement devices but require the introduction of trusted, precisely controlled, quantum sources. All entangled states can be detected in this way \cite{PhysRevLett.108.200401}. An even stronger framework is the device-independent (DI) scenario, in which all assumptions on the quantum measurement devices are removed. EWs designed for the DI scenario are commonly known as Bell inequalities, and thus verify entanglement from quantum nonlocality~\cite{PhysRevLett.106.250404, Baccari2017, PhysRevLett.114.190401}.

While the violation of EWs in these three  scenarios is sufficient to detect entanglement, a natural question is how the magnitude of the violation can be used to quantify the  entanglement in the state. When devices are trusted,  entanglement quantification from EWs has for instance been approached using the Legendre transform~\cite{doi:10.1063/1.2337535,  PhysRevA.72.022310,PhysRevLett.110.030501, Eisert_2007, PhysRevLett.98.110502}. For MDI-EWs,  entanglement quantification often comes with costly computational methods~\cite{PhysRevLett.118.150505, PhysRevA.98.052332, PhysRevLett.116.190501}.  In the DI scenario, little is known about entanglement quantification. Notable exceptions again require dedicated computational methods that grow rapidly with the complexity of the scenario~\cite{PhysRevLett.111.030501}.

%Estimating entanglement  in terms of various measures is of equal importance to detecting entanglement as which  tells how well one protocol will perform using  a given state. As entanglement measures  are defined  dramatically different in mathematics, finding an universal method to estimate them is quite challenging, especially  with  only partial information of density matrix obtained in experiment. One idea of highly  interest is using  the information gathered in the implantation of  an EW to estimate entanglement.   %EWs are  widely employed in the investigation of entanglement, and making the most of the information gathered there for a quantitative statement of entanglement  attracts much focus.
% For standard  EWs,  this idea is mainly tackled   based on an approach of  .  For MDI-EWs,  one has to employ costly numerical calculations~\cite{PhysRevLett.118.150505, PhysRevA.98.052332, PhysRevLett.116.190501} to   quantitatively connect EWs to entanglement.  For the most noise-robust DI-EWs, this topic is still open therein  the difficulty lies behind the fundamental issue  of  establishing general and quantitative connection between  entanglement measures  and nonlocality.  The connection is still missing except in some specific scenarios. This  is due to the fact  nonlocality and entanglement are different concepts having rich structures respectively and  subtle interplay, for instance, they do not always positively  relate to each other and there are even entangled but  local states.

In this work, we introduce a comprehensive method for estimating entanglement, which requires no numerical optimisation and is easy to compute. The main idea is to suitably normalise any given entanglement witness so that its violation magnitude can be used directly to bound the trace-distance between a given entangled state and its best separable approximation. In turn, this entanglement quantifier can be used to bound a variety of well-known entanglement measures from any given EW. We then extend this idea from the scenario of trusted devices to both the MDI scenario and the DI scenario, for which the bounds are typically not tight. The method is inherently analytical. For trusted devices and the MDI scenario, the quantification can be easily computed. In the DI scenario, the most demanding computation needed is an upper bound on the largest quantum violation of the considered EW. Finally, we showcase how the approach can go further, by quantifying entanglement with respect to a given Schmidt number and estimating the depth of entanglement in  multipartite systems.

\section{II. Trusted devices}
We focus first on the bipartite scenario. When devices are trusted, consider that we are given a standard EW in the form of a Hermitian operator $\mathcal{W}$~\cite{chruscinski2014entanglement, RevModPhys.81.865}. By definition, we have $w (\rho) \equiv \tr\left(\rho \mathcal{W} \right)\geq 0$ for all separable states and $w< 0$ for some entangled states, where and henceforth we omit the state  if not introducing ambiguity.   Our  goal is to bound from below the distinguishability between $\rho$ and any separable state in terms of a negative $w$, and then use the bound to estimate other entanglement measures. This entanglement measure corresponds to the smallest trace-distance between $\rho$ and any separable state whose set is specified by $\Omega$, that is,
\begin{equation}
E_{\rm tr}(\rho)=\min_{\varrho\in\Omega} D_{\rm tr}(\varrho,\rho),
\end{equation}
where  $D_{\rm tr}(\varrho,\rho)=\frac{1}{2}\tr|\varrho-\rho|$
 is the trace-distance, with ${\rm Tr}|A|={\rm Tr}\sqrt{AA^{\dagger}}$. The measure $E_{\rm tr}$ is motivated by having a natural physical interpretation: it stands in one-to-one correspondence with the largest success probability of discriminating between the state $\rho$ and any possible separable state via a quantum measurement.

To this end, we observe that
$D_{\rm tr}(\rho,\varrho)\geq \frac{1}{2}|\tr\left((\varrho-\rho)\mathcal{W}\right)|\geq -\frac{w}{2}$ if $-\openone\leq \mathcal{W}\leq \openone$.
Naturally, there are many ways of satisfying the latter condition. A simple one is to divide $\mathcal{W}$ by its largest modulus eigenvalue denoted by $\lambda\equiv \max\{|\lambda_{\pm}|\}$, where $\lambda_{\pm}$ are respectively, the largest and smallest eigenvalues of $\mathcal{W}$. However, an often better choice is to consider the following Hermitian operator,
\begin{equation}\label{step1}
\mathcal{W}' =\frac{2\mathcal{W}-(\lambda_{+}+\lambda_{-})\openone}{\lambda_{+}-\lambda_{-}}\equiv2\mathcal{W}_{c}-\frac{(\lambda_{+}+\lambda_{-})\openone}{\lambda_{+}-\lambda_{-}},
\end{equation}
where the non-trivial part is denoted by    $2\mathcal{W}_{c}\equiv \frac{2\mathcal{W}}{\lambda_{+}-\lambda_{-}}$, and a smaller involved numerator $\frac{\lambda_{+}-\lambda_{-}}{2}$ than $\lambda$  would  ensure a larger lower bound on $E_{\rm tr}$.  It  then follows that
\begin{Theorem}[Quantitative Normalized  EW] For an EW $\mathcal W$ that detects an entangled state $\rho$ by a negative witness value, it holds
\begin{align}\label{trtrust}
E_{\rm tr}(\rho)\geq -w_c,
\end{align}
where $w_c\equiv \tr\left(\mathcal{W}_{c} \rho\right)$,  and  we have this theorem as $E_{\rm tr}(\rho)=\textstyle  \frac{1}{2}\tr|\varrho_\text{opt}-\rho| \geq \textstyle \frac{1}{2}|\tr[(\varrho_{\text{opt}}-\rho)\mathcal{W}']|= |\tr[(\varrho_{\text{opt}}-\rho) \mathcal{W}_{c}]| \geq -w_c$ with $\varrho_{\text{opt}}$ denoting  the (closest) separable state thus  $\tr\left(\mathcal{W}_{c}\varrho_\text{opt}\right)\geq 0$,.
\end{Theorem}
Thus, we can estimate $E_{\rm tr}(\rho)$ directly from the re-normalised entanglement witness   $\mathcal{W}_c$.
In addition, we can use $E_{\rm tr}(\rho)$ itself to bound several other well-known    entanglement quantifiers.

Focusing on other distance-based entanglement measures~\cite{PhysRevLett.78.2275}, prominent examples are the smallest relative entropy, $E_{\rm re}$, or infidelity, $E_{\rm if}$, between an entangled state and a separable state \cite{PhysRevLett.78.2275}. Other relevant examples are convex roof measures, such as the entanglement of formation \cite{PhysRevA.54.3824}, $\tilde{E}_{\rm F}$, the concurrence \cite{PhysRevLett.80.2245}, $\tilde{E}_{\rm C}$, and the geometric measure of entanglement \cite{PhysRevA.68.042307}, $\tilde{E}_{\rm G}$. Also relevant are the robustness, $E_{\rm rob}$, and generalised robustness, $E_{\rm ROB}$, of entanglement \cite{PhysRevA.59.141}. The definitions of these measures are summarised in~Appendix.A.  In Appendix.B we show how all these quantifiers of entanglement can be bounded in terms of $E_{\rm tr}$. Hence, they can be bounded directly from the observed value of the re-normalised EW as
\begin{align}\nonumber\label{EWbounds}
&E_{\rm re}, \tilde {E}_{\rm F}\geq \frac{2}{\ln2}w_c^{2},  && E_{\rm if}, \tilde {E}_{\rm G}\geq w_c^{2},  \\
&  \tilde E_{\rm C}\geq -\sqrt{2}w_c, && E_{\rm rob, ROB}\geq \frac{-w_c}{1+w_c}.
\end{align}
Note that these relations, as well as Eq.~\eqref{trtrust}, still remain if we replace the set of separable states with other relevant convex sets on which we can define analogous witness operators. A particularly interesting example is the set of states with a limited Schmidt number~\cite{Terhal2000b}. This would permit the quantification of an entangled state with respect to any other entangled state of a limited entanglement dimension.

We exemplify the result on the well-known fidelity  witness~\cite{PhysRevLett.92.087902, PhysRevA.60.898, PhysRevA.59.4206} assuming the form $\mathcal{W}=c\hat{\openone}-|\phi\rangle \langle \phi|$, where $c\equiv \max_{\varrho \in \Omega} {\rm Tr}(\varrho |\phi\rangle \langle \phi|)$.  We have  $\lambda_-=c-1$ and  $\lambda_+=c$ and $\lambda_{+}-\lambda_-=1$  hence $\mathcal{W}_c=\mathcal{W}$. The magnitude of a negative witness value immediately gives a lower bound on $E_{\rm tr}$.  For example, let $|\phi\rangle $ be the maximum entangled state in $d$-dimensional space, namely,  $\rho=|\phi\rangle\langle \phi|$ with $|\phi\rangle=\frac{\sum_{i}|ii\rangle}{\sqrt{d}}$.  We have $c=\frac{1}{d}$, the related  witness reads $\mathcal{W}=\frac{\openone}{d} -|\phi\rangle\langle\phi|$. This leads to  the tight bound $E_{\rm tr}(\phi)\ge1-\frac{1}{d}$. To investigate the accuracy of the bounds on the other entanglement measures, we have computed the ratios, $R$, between the bounds  \eqref{EWbounds} and the exact values. We find $R_{\rm E_{\rm tr}}=R_{\rm E_{\rm rob, ROB}}=1$,  $R_{\rm E_{\rm re}, \tilde E_{\rm  F}}=\frac{2(d-1)^{2}}{d^{2}\ln{2}\log_{2} d}$,   $R_{\rm E_{if}, \tilde E_{G}}=1$,   $R_{\rm \tilde  E_{\rm C}}=\sqrt{1-\frac{1}{d}}$. Thus, some cases are tight (when the  ratios $R=1$) while others are proper lower bounds.  Such an estimation immediately applies to  the data from  current experiments, for example, let $|\phi\rangle=\frac{1}{4}(|0000\rangle -|1001\rangle-|0110\rangle -|1111\rangle )$ be a four-qubit cluster state, for which  $c=\frac{1}{2}$. With a measured fidelity between $|\phi\rangle $  and a concerned $\rho$ as ${\rm Tr}(\rho |\phi\rangle \langle \phi|)=0.85$~\cite{PhysRevLett.112.100403}, one immediately obtains a lower bound on $E_{\rm tr}(\rho)\geq 0.35$.

%\zp{Shall we consider the suggestion from Julio, that results related to witness and some well-know measures exist already even in the general resource theory?}

\section{III. Measurement-device-independent scenario}
Every EW in the standard (or trusted) scenario can be converted into an EW in the MDI scenario \cite{PhysRevLett.108.200401, PhysRevLett.110.060405}. Specifically, the original EW can always be decomposed into the form $\mathcal{W}=\sum_{s,t} \alpha_{s,t}\tau^{\rm T}_{s}\otimes \omega^{\rm T}_{t}$, with $\rm T$ denoting the transpose and $\alpha_{s, t}$ being some real coefficients for some states  $\tau_{s}$ and $\omega_{t}$. In the MDI scenario, two parties, Alice and Bob, receive one share of $\rho$ each and then individually prepare  ancillary quantum states  $\tau_{s}$ and $\omega_{t}$, according to privately selected classical inputs $s$ and $t$, respectively.  Each of them    performs an uncharacterised measurement $\{A_a\}$ and $\{B_{b}\}$ on the joint system-ancilla state. The correlations become $p(a,b|s,t)=\tr[(A_{a}\otimes B_{b})(\tau_{s}\otimes\rho_{AB}\otimes \omega_{t})]$.

Note that $\sum_{s,t }\alpha_{s,t} p(a,b|s,t)= \tr[ (A_{a}\otimes B_{b}) (\mathcal{W}^{\rm T}\otimes \rho_{AB}) ]$,  as shown in \cite{PhysRevLett.110.060405}, if the first outcome in each measurement corresponds to a projection onto the maximally entangled state, the value of the MDI-EW
$I_{\mathcal{W}}(\rho_{AB})=\sum_{s,t }\alpha_{s,t}\ p(1,1|s,t)$
reduces to the value of the original EW, up to a positive constant.
Hence, it follows that for any separable state, $I_{\mathcal{W}}(\rho_{AB})\geq0$, whereas $I_{\mathcal{W}}(\rho_{AB})<0$ implies entanglement.  Thus, at the cost of introducing trusted sources for preparing $\tau_{s}$ and $\omega_{t}$, entanglement can now be certified with uncharacterized measurements.

This initial MDI protocol,  collecting the correlation for   the first outcome in each side,  is highly inefficient for estimating  entanglement. We slightly  optimize the protocol   by collecting all pairs of outcomes $(a, b)$ and  calculate the quantity   $ \sum_{s,t }\alpha_{s,t} p(a,b|s,t)\equiv w_{a, b}(\rho_{AB})$ for each of them (the original MDI-EW $I_{\mathcal{W}}=w_{1, 1}$ ). As one is free to label outcomes of measurements, if there exists one pair  of  outcome  $a, b$ such that $w_{a, b}< 0$. One can relabel $(a,b)$ as $(1,1)$, and  entanglement is verified  according to the original MDI-EW.  By summing all the $w_{a, b}< 0$,  we define a MDI witness value  as
 \begin{align}\nonumber
 \textstyle I'_{\mathcal{W}}(\rho_{AB})=\sum_{a, b| w_{a, b}<0} w_{a, b}(\rho_{AB}).
\end{align}
We remark that
$I'_{\mathcal{W}}$ is non-linear since the picked out  $w_{a, b}$s    depend on the state $\rho$.
It is clear that, for any separable state,  $w_{a, b}(\rho_{AB})\geq0$ and $\textstyle I'_{\mathcal{W}}(\rho_{AB})=0$.

We now show that the witness value $I'_{\mathcal{W}}(\rho_{AB})$ can be used to  bound the entanglement measure $E_{\rm Tr}(\rho)$. To this end, we observe that, for two  Hermitian matrices $M$ and $N$,  ${\rm Tr}|M|\cdot{\rm Tr}|N|={\rm Tr}|M\otimes N|$. Since ${\rm Tr}|\varrho_{\text{opt}}-\rho|=\textstyle \frac{1}{{\rm Tr}|\mathcal{W}^{\rm T}|} {\rm Tr}|\mathcal{W}^{\rm T}\otimes (\varrho_{\text{opt}}-\rho)|$, then we have
\begin{align}\nonumber
2{\rm Tr}|\mathcal{W}^{\rm T}| E_{\rm Tr}(\rho)&=\textstyle {\rm tr}|\mathcal{W}^{\rm T}\otimes (\varrho_{\text{opt}}-\rho)| \nonumber\\
& \geq \textstyle\sum_{a,b} |{\rm Tr}[\mathcal{W}^{\rm T}\otimes (\varrho_{\text{opt}}-\rho)\cdot(A_a\otimes B_b)]|\nonumber\\
&\geq  \textstyle  |\sum_{a, b| w_{a,b}(\rho)<0}(w_{a,b}(\rho)-  w_{a,b}(\varrho_{\text opt}))| \nonumber\\
  & \qquad +\textstyle  |\sum_{a, b| w_{a,b}(\rho)\geq 0}(w_{a,b}(\rho)-  w_{a,b}(\varrho_{\text opt}))|\nonumber \\
& =  \textstyle \sum_{a, b| w_{a,b}(\rho)<0}2 [w_{a,b}(\varrho_{\text opt})- w_{a,b}(\rho)]\nonumber \\
& \geq-2I'_{\mathcal{W}}(\rho), \label{mdi}
\end{align}
%\zp{[Notice that $\{A_a\}$ is a POVM, hence the second inequality in the third line might not hold.]}
where  $\varrho_{\text{opt}}$ denotes  the closest separable state, and in the second equality  we have used $\sum_{a, b} w_{a,b}(\rho)=\sum_{a, b} w_{a,b}(\varrho_{\rm opt})={\rm Tr}(\mathcal{W}^{\rm T})$ which implies that  $ \sum_{a, b| w_{a,b}(\rho)<0} [w_{a,b}(\rho)-  w_{a,b}(\varrho_{\text opt})]
  =\textstyle \sum_{a, b| w_{a,b}(\rho)\geq 0} [w_{a,b}(\rho)-  w_{a,b}(\varrho_{\text opt})]$.
 Then we arrive at
 \begin{align}%\nonumber
 E_{\rm tr}(\rho)\geq -\frac{I'_{\mathcal{W}}(\rho)}{ {\rm Tr}|\mathcal{W}^{\rm T}|}.
\end{align}

 We consider   the 2-qubit Werner state $(1-v)\openone/4 +v|\phi^{-}\rangle\langle\phi^{-}|$ with $(0\le v\le 1$) detected with EW $\mathcal {W}=\openone/2-|\phi^{-}\rangle\langle \phi^{-}|$ where $|\phi^{-}\rangle=1/\sqrt{2} (|01\rangle -|10\rangle ) $, for which, ${\rm Tr}|\mathcal{W}^{\rm T}|=2$.  Here  we employ the Bell state measurement  $\{|\phi^{\pm}\rangle\langle \phi^{\pm}|,|\psi^{\pm}\rangle\langle \psi^{\pm} \}|  \}$, where $|\phi^{\pm}\rangle=1/\sqrt{2} (|01\rangle \pm |10\rangle )$ and  $|\psi^{\pm}\rangle=1/\sqrt{2} (|00\rangle \pm |11\rangle )$. The state is entangled when $v> 1/3$, then we have $ w_{a,b}(\rho)=\frac{1-3v}{16}<0$ when $a=b$, otherwise,     $ w_{a,b}(\rho)=\frac{1+v}{16}> 0$.  Thus we obtain an estimation $E_{\rm tr}(\rho_v)\ge -\frac{1-3v}{8}$, which is a half of the real value when $v=1$. Estimations for other measures can be done  via  Eq.(\ref{EWbounds}).

\section{IV. Device-independent scenario}
In the DI scenario, we make no assumptions on Alice's and Bob's devices. Consider $n$ parties, called $\{\mathcal{A}_{1}, \cdots \mathcal{A}_{n}\}$.  Party $\mathcal{A}_{i}$ privately selects an input $s_{i}$ from a finite classical alphabet and performs a measurement $\{\mathcal{M}_{a_i|s_i}\}$ which has outcome  $a_{i}$. The probability of obtaining outcomes $\mathbf{a}=\{a_{1}, \cdots, a_{n} \}$ when the inputs are $\mathbf{s}=\{s_1, \cdots, s_n\}$ is denoted   $p(\mathbf{a}| \mathbf{s})$. A DI-EW, also known as a Bell inequality, when appropriately normalised for our purposes, takes the form
\begin{equation}
\tr\left(\mathcal{W}_\beta\rho\right)< 0,
\end{equation}
where $\mathcal{W}_\beta=\beta_c\openone -\mathcal{B}$, and $\mathcal{B}=\sum_{\mathbf{a},  \mathbf{s}}a_{\mathbf{a}, \mathbf{s}} \hat{\mathcal{M}}_{a_{1}|s_1}\otimes\cdots \otimes\hat{\mathcal{M}}_{a_{n}|s_n}$ is Bell's quantity, and $\beta_c$ is the maximum value of Bell's quantity over separable states and also the local-hidden-variable bound relevant to $\mathcal{B}$.   We can now use  the EW $\mathcal{W}_\beta$ to estimate DI entanglement. To this end, following an argument that closely parallels that outlined for trusted devices in Eqs.~\eqref{step1} and \eqref{trtrust} leads to
\begin{eqnarray}
E_{\rm tr}(\rho) \geq  -{\rm Tr}\left(\rho \mathcal{W}_{\rm cDI}\right),
\end{eqnarray}
where we have defined the re-normalised EW as
\begin{equation}
\mathcal{W}_{\rm cDI}=\frac{\mathcal{W}_\beta}{\langle \mathcal{B}\rangle_{+}  -\langle \mathcal{B}\rangle_{-}},
\end{equation}
where $\langle \mathcal{B}\rangle_{\pm}$ is the largest and smallest eigenvalues of the Bell operator $\mathcal{B}$ when evaluated over all possible measurements. In other words, it is the largest and smallest Bell values in quantum theory.  In the case that these so-called Tsirelson bounds cannot be determined analytically, they can still be efficiently bounded by means of semi-definite programming \cite{PhysRevLett.98.010401}.

As an illustration, let us consider the paradigmatic  Clauser-Horne-Shimony-Holt inequality~\cite{PhysRevLett.23.880}. Its Bell operator, expressed in the two respective observables of Alice and Bob, reads  $\mathcal B_{\rm chsh}=A_{0}B_{0}+A_{0}B_{1}+A_{1}B_{0}-A_{1}B_{1}$, for which $\beta_{c}=2$. Our DI-EW becomes  $\mathcal{W}_{\rm chsh}=2\openone -\mathcal{B}_{\rm chsh}$. Since the Tsirelson bounds $\langle \mathcal{B}\rangle_{+} = -\langle \mathcal{B}\rangle_{-} = 2\sqrt{2}$, we have  $\mathcal{W}_{\rm cDI-chsh}=\frac1{4\sqrt{2}}\mathcal{W}_{\rm chsh}$. Hence, the estimated entanglement is at least
 \begin{eqnarray}
E_{\rm tr}(\rho)\geq \frac{\langle \mathcal B_{\rm chsh}\rangle_\rho-2}{4\sqrt{2}}. \label{dichsh}
\end{eqnarray}
This  lower bound  is unfortunately not tight. For example,  when $\langle \mathcal B_{\rm chsh}\rangle_\rho=2\sqrt{2}$, the lower bound is $\frac{1}{2}-\frac{\sqrt{2}}{4}$, instead of the true value $\frac{1}{2}$.

To improve our bound, we can first rewrite a DI-EW as $\frac{\max_{\varrho\in \Omega}\langle \mathcal{B}\rangle_{\varrho}-\langle\mathcal{B}\rangle_{\rho} }{\langle \mathcal{B}\rangle_{+}-\langle \mathcal{B}\rangle_{-}}$.  A maximal CHSH violation, $\langle \mathcal B_{\rm chsh}\rangle_\rho=2\sqrt{2}$, is well-known to self-test anti-communting qubit measurements for both Alice and Bob. For such measurements, there are well-known  uncertainty relations  ${\langle A_{0}}\rangle^{2}+{\langle A_{1}}\rangle ^{2}\leq 1 $ and ${\langle B_{0}}\rangle^{2}+{\langle B_{1}}\rangle ^{2}\leq 1 $. These relations imply a bound on $\max_{\varrho \in \Omega} \langle \mathcal{B}\rangle_{\varrho}$. Specifically, for a product state  $\varrho=\varrho_{A}\otimes \varrho_{B}$, one has  the correlations  as  $\langle A_{i}B_{j}\rangle_{\varrho}=\langle {A_{i}}\rangle_{\varrho_{A}}\langle{B_{j}}\rangle_{\varrho_{B}}$ and $\langle \mathcal B_{\rm chsh}\rangle_\varrho=\sum_{i,j=0,1} (-1)^{ij}\langle {A_{i}}\rangle_{\varrho_{A}}\langle{B_{j}}\rangle_{\varrho_{B}} $. The uncertainty relations then imply    $\max_{\varrho\in \Omega}\langle \mathcal B_{\rm chsh}\rangle_\varrho=\sqrt{2}$, which  is strictly less than the classical bound $\beta_{c}=2$ involved in Eq.(\ref{dichsh}), thus improving the lower bound in  Eq.(\ref{dichsh}) from $\frac{1}{2}-\frac{\sqrt{2}}{4}$ to  $\frac{1}{4}$.
%\armin{This analysis is only for the maximal violation but it seems intuitive that it should work also for non-maximal violations. Why not include that? Look e.g. at eq3 in arXiv:1702.06845}

%\textcolor{blue}{ We have successfully tackled the problem by developing a new idea. That is, in Bell's scenario, it is possible to certify entanglements even with local correlations if the measurements are known. For instance,in case the measurements are maximally incompatible, the maximum Bell's value for a separables tate is $\sqrt{2}$, which is less than 2. This implies that certain local correlations between the two values  can only be achieved with some entangled states. This idea rests on complex mathematics£¬ which maybe bring a burden to  the reader so not attached in this paper. }

\section{V. Estimating entanglement of $k-$depth}
We apply our approach also to quantify the depth of entanglement in multipartite systems. A multipartite pure state $|\Psi\rangle$ is said to be \emph{k}-producible if the subsystems can be partitioned into   $m$ pairwise disjoint and  non-empty subsets, denoted by $\{\mathcal{A}_{1}, \cdots \mathcal{A}_{m}\}$, such that,  state $|\Psi\rangle$   can be written as  a  product  as $|\psi_{1}\rangle_{\mathcal{A}_{1}}\otimes \cdots \otimes|\psi_{m}\rangle_{\mathcal{A}_{m}}$ and  each set $\mathcal{A}_{i}$ contains at $k$ elements. Similarly, a mixed state is called $k-$producible if it can be   decomposed into  a mixture of $k-$producible pure states. If a state is not  $k-$producible, it is said that its entanglement depth is at least $k+1$.   We write $\mathcal{W}_{\mathcal{P}_{k}}$ for a  witness of  $k-$depth entanglement; it is an observable such that the expectation value $w_{{\rm c},\mathcal{P}_{k}}\equiv {\rm Tr}(\mathcal{W}_{\mathcal{P}_{k}}\rho)\geq 0$. If  $\rho$ is  $k-$producible state and $w_{{\rm c},\mathcal{P}_{k}}< 0$ for some  state with  entanglement depth more than $k$.

Let $\Omega_{\mathcal{P}_{k}}$ be the set of all states with entanglement depth at most $k$. In analogy with the bipartite case, we consider the distinguishability between a given state $\rho$, and any state in $\Omega_{\mathcal{P}_{k}}$, namely
\begin{eqnarray}
E_{\rm tr; \mathcal{P}_{k}}(\rho)\equiv\min_{\varrho\in \Omega_{\mathcal{P}_{k}}}D_{\operatorname{tr}}(\rho, \varrho)\geq -w_{{\rm c},\mathcal{P}_{k}},
\end{eqnarray}
which can be estimated immediately with some known multipartite entanglement witnesses.  Two cases are provided in what follows.

We exemplify this for a standard entanglement witness tailored for a noisy tripartite W-state $
 \rho= \frac{v \openone }{8}+(1-v)|\Psi\rangle\langle \Psi|$ where
$|\Psi\rangle=\frac{1}{\sqrt{3}}(|1,0,0 \rangle +|0,1, 0 \rangle +|0,0, 1 \rangle)$.  Standard multipartite EWs  can be constructed by solving the so-called  multipartite separability eigenvalue equations~\cite{PhysRevLett.111.110503, PhysRevX.8.031047}. In Ref~\cite{PhysRevLett.111.110503} examples of EWs of  $k-$producible  entanglement for this state are given as $\mathcal{W}_{\mathcal{P}_{1}}=\frac{4}{9}\cdot \openone- |\Psi\rangle\langle \Psi|$,  and $\mathcal{W}_{\mathcal{P}_{2}}=\frac{2}{3}\cdot \openone- |\Psi\rangle\langle \Psi|$. For both of them,  $\lambda_+-\lambda_-=1$.
 We  have $w_{{\rm c},\mathcal{P}_{1}}=\frac{4}{9}-\frac{7v}{8}$ and $w_{{\rm c},\mathcal{P}_{2}}=\frac{2}{3}-\frac{7v}{8}$.  Entanglement is certified when $v\geq \frac{40}{63}$ as $w_{{\rm c},\mathcal{P}_{1}}<0$,  and  entanglement of depth three is certified when  $v>\frac{8}{21}$. From our approach, we get a  also quantitative statement.  For instance when $v=1$, we have $E_{\rm tr; \mathcal{P}_{1}}(\rho)\geq \frac{31}{72}$ and $E_{\rm tr; \mathcal{P}_{2}}(\rho)\geq \frac{5}{24}$.

Next, we consider the multipartite scenario in a device-independent setting. Multipartite DI-EWs are designed  by exploiting the fact that quantum state with a deeper entanglement depth  can violate some Bell's inequalities to a larger extent~\cite{PhysRevD.35.3066, PhysRevLett.106.250404, PhysRevLett.88.170405, PhysRevLett.106.020405,PhysRevLett.114.190401, PhysRevLett.90.080401}. One typical such Bell's inequality  is the  Svetlichny inequality that reads~\cite{PhysRevD.35.3066, Ba2013}
\begin{equation}
\mathcal{B}^{(n)}_{S}=2^{-\frac{n}{2}}\sum_{\bf a,\mathbf{s}}(-1)^{a+\lfloor\frac{s}{2}\rfloor}\hat{\mathcal{M}}_{a_{1}|s_1}\otimes\cdots \otimes\hat{\mathcal{M}}_{a_{n}|s_n},
\end{equation}
where $s=\sum_{i}s_{i}$ and $a=\sum_ia_i$ are the sum of measurement setting $s_{i}\in\{0, 1\}$ and outcome $a_i\in\{0,1\}$.
The maximum expectation value of $\mathcal{B}^{(n)}_{S}$ with respect to the set of $k-$producible  states   is~\cite{PhysRevA.100.022121}  $\beta_{k}\equiv 2^{(n-2\lfloor\frac{\lceil n/k\rceil}{2}\rfloor)/2}$, and the maximum and minimum values for an arbitrary quantum state  is $\langle \mathcal{B}^{(n)}_{S} \rangle_{+}=-\langle \mathcal{B}^{(n)}_{S} \rangle_{-}=2^{(n-1)/2}$.
Thus, the re-normalised multipartite DI-EW of $k$-producible states reads
\begin{equation}
\mathcal{W}^{(n)}_{{\rm cDI};\mathcal{P}_{k}}:=
\frac{\mathcal{W}^{n}_{{\rm DI}, \mathcal{P}_{k}}}{2^{(n+1)/2}},\quad \mathcal{W}^{n}_{{\rm DI}, \mathcal{P}_{k}}:=\beta_{k}\cdot \openone-\mathcal{B}^{(n)}_{S},
\end{equation}
from which we  have the quantitative estimation
\begin{equation}
\textstyle E^{(n)}_{{\rm tr};\mathcal{P}_{k}}\geq -w^{(n)}_{{\rm cDI};\mathcal{P}_{k}}=-\tr(\rho \mathcal{W}^{(n)}_{{\rm cDI};\mathcal{P}_{k}}). \label{npartite}
\end{equation}
  As a simple illustration of Eq.(\ref{npartite}), consider the correlation scoring $\langle \mathcal{B}^{(n)}_{S} \rangle=\langle \mathcal{B}^{(n)}_{S} \rangle_{+}$, which is attained by the GHZ state. The lower bound for  the entanglement quantifier with respect to states of depth more than $k$ becomes
 \begin{equation}
\textstyle E^{(n)}_{{\rm tr};\mathcal{P}_{k}}\geq\frac{1}{2}(1-2^{-2\lfloor\frac{\lceil n/k\rceil}{2}\rfloor}). \label{asy1}
\end{equation}
 This lower bound asymptotically tends to the exact value for the GHZ state as  $n/k\rightarrow \infty$  and   tends to zero as $ n/k\rightarrow 1$.

In conclusion, we have provided a simple method to estimate entanglement in the standard, MDI and DI scenarios by re-normalising witness operators. The method is tailored to the operational distinguishability measure, but it also yields non-trivial bounds on many of the well-known entanglement measures used in quantum information.
This permits an experimenter to not only detect different forms of entanglement but also to quantify it, without requiring any additional measurements than those used in a standard witness-based detection scheme. Hence, this may be seen as an enhanced data analysis, which is also practically straightforward to use since our procedure requires essentially no optimisation but only simple analytical expressions. %These features make it compelling in estimating entanglement in both theory and experiment.
%Our method also has another advantage of  involving direct mathematics only (no optimization is involved), which makes it quite suitable for analyzing  the multipartite entanglement in a quantitative manner.
Finally, we note that our approach might be interesting to consider also in the context of estimating quantum properties in other scenarios, such as steering, quantum coherence and entanglement-assisted quantum communication.

\begin{acknowledgements}
L.L.S. and S.Y. would like to thank Key-Area Research and Development Program of Guangdong Province Grant No. 2020B0303010001.	Z.P.X. acknowledges support from {National Natural Science Foundation of China} (Grant No.\ 12305007),
Anhui Provincial Natural Science Foundation (Grant No.\ 2308085QA29). A.T. 	is  supported by the Wenner-Gren Foundation and  by the Knut and Alice Wallenberg Foundation through the Wallenberg Center for Quantum Technology (WACQT).

\end{acknowledgements}
\bibliography{11}

\begin{thebibliography}{57}
\expandafter\ifx\csname natexlab\endcsname\relax\def\natexlab#1{#1}\fi
\expandafter\ifx\csname bibnamefont\endcsname\relax
  \def\bibnamefont#1{#1}\fi
\expandafter\ifx\csname bibfnamefont\endcsname\relax
  \def\bibfnamefont#1{#1}\fi
\expandafter\ifx\csname citenamefont\endcsname\relax
  \def\citenamefont#1{#1}\fi
\expandafter\ifx\csname url\endcsname\relax
  \def\url#1{\texttt{#1}}\fi
\expandafter\ifx\csname urlprefix\endcsname\relax\def\urlprefix{URL }\fi
\providecommand{\bibinfo}[2]{#2}
\providecommand{\eprint}[2][]{\url{#2}}

\bibitem[{\citenamefont{Horodecki
  et~al.}(2009{\natexlab{a}})\citenamefont{Horodecki, Horodecki, Horodecki, and
  Horodecki}}]{RevModPhys.81.865}
\bibinfo{author}{\bibfnamefont{R.}~\bibnamefont{Horodecki}},
  \bibinfo{author}{\bibfnamefont{P.}~\bibnamefont{Horodecki}},
  \bibinfo{author}{\bibfnamefont{M.}~\bibnamefont{Horodecki}},
  \bibnamefont{and}
  \bibinfo{author}{\bibfnamefont{K.}~\bibnamefont{Horodecki}},
  \bibinfo{journal}{Rev. Mod. Phys.} \textbf{\bibinfo{volume}{81}},
  \bibinfo{pages}{865} (\bibinfo{year}{2009}{\natexlab{a}}),
  \urlprefix\url{https://link.aps.org/doi/10.1103/RevModPhys.81.865}.

\bibitem[{\citenamefont{Briegel et~al.}(2009)\citenamefont{Briegel, Browne,
  D\"{u}r, Raussendorf, and den Nest}}]{Briegel2009}
\bibinfo{author}{\bibfnamefont{H.~J.} \bibnamefont{Briegel}},
  \bibinfo{author}{\bibfnamefont{D.~E.} \bibnamefont{Browne}},
  \bibinfo{author}{\bibfnamefont{W.}~\bibnamefont{D\"{u}r}},
  \bibinfo{author}{\bibfnamefont{R.}~\bibnamefont{Raussendorf}},
  \bibnamefont{and} \bibinfo{author}{\bibfnamefont{M.~V.} \bibnamefont{den
  Nest}}, \bibinfo{journal}{Nat. Phys} \textbf{\bibinfo{volume}{5}},
  \bibinfo{pages}{19} (\bibinfo{year}{2009}),
  \urlprefix\url{https://doi.org/10.1038%2Fnphys1157}.

\bibitem[{\citenamefont{Ekert}(1991)}]{PhysRevLett.67.661}
\bibinfo{author}{\bibfnamefont{A.~K.} \bibnamefont{Ekert}},
  \bibinfo{journal}{Phys. Rev. Lett.} \textbf{\bibinfo{volume}{67}},
  \bibinfo{pages}{661} (\bibinfo{year}{1991}),
  \urlprefix\url{https://link.aps.org/doi/10.1103/PhysRevLett.67.661}.

\bibitem[{\citenamefont{Giovannetti et~al.}(2004)\citenamefont{Giovannetti,
  Lloyd, and Maccone}}]{Giovannetti2004}
\bibinfo{author}{\bibfnamefont{V.}~\bibnamefont{Giovannetti}},
  \bibinfo{author}{\bibfnamefont{S.}~\bibnamefont{Lloyd}}, \bibnamefont{and}
  \bibinfo{author}{\bibfnamefont{L.}~\bibnamefont{Maccone}},
  \bibinfo{journal}{Science} \textbf{\bibinfo{volume}{306}},
  \bibinfo{pages}{1330} (\bibinfo{year}{2004}),
  \urlprefix\url{https://doi.org/10.1126%2Fscience.1104149}.

\bibitem[{\citenamefont{Gurvits}(2004)}]{np2}
\bibinfo{author}{\bibfnamefont{L.}~\bibnamefont{Gurvits}}, \bibinfo{journal}{J.
  Comput. Syst. Sci.} \textbf{\bibinfo{volume}{69}}, \bibinfo{pages}{448}
  (\bibinfo{year}{2004}), ISSN \bibinfo{issn}{0022-0000},
  \bibinfo{note}{special Issue on STOC 2003}.

\bibitem[{\citenamefont{Gurvits}(2003)}]{np1}
\bibinfo{author}{\bibfnamefont{L.}~\bibnamefont{Gurvits}},
  \emph{\bibinfo{title}{Classical deterministic complexity of edmonds' problem
  and quantum entanglement}} (\bibinfo{year}{2003}), \eprint{quant-ph/0303055}.

\bibitem[{\citenamefont{Huang}(2014)}]{Huang_2014}
\bibinfo{author}{\bibfnamefont{Y.}~\bibnamefont{Huang}}, \bibinfo{journal}{New
  J. Phys.} \textbf{\bibinfo{volume}{16}}, \bibinfo{pages}{033027}
  (\bibinfo{year}{2014}),
  \urlprefix\url{https://dx.doi.org/10.1088/1367-2630/16/3/033027}.

\bibitem[{\citenamefont{Horodecki
  et~al.}(2009{\natexlab{b}})\citenamefont{Horodecki, Horodecki, Horodecki, and
  Horodecki}}]{Horodecki2009}
\bibinfo{author}{\bibfnamefont{R.}~\bibnamefont{Horodecki}},
  \bibinfo{author}{\bibfnamefont{P.}~\bibnamefont{Horodecki}},
  \bibinfo{author}{\bibfnamefont{M.}~\bibnamefont{Horodecki}},
  \bibnamefont{and}
  \bibinfo{author}{\bibfnamefont{K.}~\bibnamefont{Horodecki}},
  \bibinfo{journal}{Rev. Mod. Phys.} \textbf{\bibinfo{volume}{81}},
  \bibinfo{pages}{865} (\bibinfo{year}{2009}{\natexlab{b}}),
  \urlprefix\url{https://link.aps.org/doi/10.1103/RevModPhys.81.865}.

\bibitem[{\citenamefont{Friis et~al.}(2019)\citenamefont{Friis, Vitagliano,
  Malik, and Huber}}]{nrev}
\bibinfo{author}{\bibfnamefont{N.}~\bibnamefont{Friis}},
  \bibinfo{author}{\bibfnamefont{G.}~\bibnamefont{Vitagliano}},
  \bibinfo{author}{\bibfnamefont{M.}~\bibnamefont{Malik}}, \bibnamefont{and}
  \bibinfo{author}{\bibfnamefont{M.}~\bibnamefont{Huber}},
  \bibinfo{journal}{Nat. Rev. Phys.} \textbf{\bibinfo{volume}{1}},
  \bibinfo{pages}{72} (\bibinfo{year}{2019}),
  \urlprefix\url{https://doi.org/10.1038/s42254-018-0003-5}.

\bibitem[{\citenamefont{Tavakoli et~al.}(2023)\citenamefont{Tavakoli,
  Pozas-Kerstjens, Brown, and Araújo}}]{SDPreview}
\bibinfo{author}{\bibfnamefont{A.}~\bibnamefont{Tavakoli}},
  \bibinfo{author}{\bibfnamefont{A.}~\bibnamefont{Pozas-Kerstjens}},
  \bibinfo{author}{\bibfnamefont{P.}~\bibnamefont{Brown}}, \bibnamefont{and}
  \bibinfo{author}{\bibfnamefont{M.}~\bibnamefont{Araújo}},
  \emph{\bibinfo{title}{Semidefinite programming relaxations for quantum
  correlations}} (\bibinfo{year}{2023}), \eprint{2307.02551}.

\bibitem[{\citenamefont{T\'oth and G\"uhne}(2005)}]{Guhne2005}
\bibinfo{author}{\bibfnamefont{G.}~\bibnamefont{T\'oth}} \bibnamefont{and}
  \bibinfo{author}{\bibfnamefont{O.}~\bibnamefont{G\"uhne}},
  \bibinfo{journal}{Phys. Rev. Lett.} \textbf{\bibinfo{volume}{94}},
  \bibinfo{pages}{060501} (\bibinfo{year}{2005}),
  \urlprefix\url{https://link.aps.org/doi/10.1103/PhysRevLett.94.060501}.

\bibitem[{\citenamefont{Morelli et~al.}(2023)\citenamefont{Morelli, Huber, and
  Tavakoli}}]{Morelli2023}
\bibinfo{author}{\bibfnamefont{S.}~\bibnamefont{Morelli}},
  \bibinfo{author}{\bibfnamefont{M.}~\bibnamefont{Huber}}, \bibnamefont{and}
  \bibinfo{author}{\bibfnamefont{A.}~\bibnamefont{Tavakoli}},
  \emph{\bibinfo{title}{Resource-efficient high-dimensional entanglement
  detection via symmetric projections}} (\bibinfo{year}{2023}),
  \eprint{2304.04274}.

\bibitem[{\citenamefont{Liu et~al.}(2022)\citenamefont{Liu, Liu, Chen, and
  Ma}}]{Liu2022}
\bibinfo{author}{\bibfnamefont{P.}~\bibnamefont{Liu}},
  \bibinfo{author}{\bibfnamefont{Z.}~\bibnamefont{Liu}},
  \bibinfo{author}{\bibfnamefont{S.}~\bibnamefont{Chen}}, \bibnamefont{and}
  \bibinfo{author}{\bibfnamefont{X.}~\bibnamefont{Ma}}, \bibinfo{journal}{Phys.
  Rev. Lett.} \textbf{\bibinfo{volume}{129}}, \bibinfo{pages}{230503}
  (\bibinfo{year}{2022}),
  \urlprefix\url{https://link.aps.org/doi/10.1103/PhysRevLett.129.230503}.

\bibitem[{\citenamefont{Terhal}(2000)}]{Terhal2000}
\bibinfo{author}{\bibfnamefont{B.~M.} \bibnamefont{Terhal}},
  \bibinfo{journal}{Phys. Lett. A} \textbf{\bibinfo{volume}{271}},
  \bibinfo{pages}{319} (\bibinfo{year}{2000}),
  \urlprefix\url{https://doi.org/10.1016%2Fs0375-9601%2800%2900401-1}.

\bibitem[{\citenamefont{Doherty et~al.}(2002)\citenamefont{Doherty, Parrilo,
  and Spedalieri}}]{PhysRevLett.88.187904}
\bibinfo{author}{\bibfnamefont{A.~C.} \bibnamefont{Doherty}},
  \bibinfo{author}{\bibfnamefont{P.~A.} \bibnamefont{Parrilo}},
  \bibnamefont{and} \bibinfo{author}{\bibfnamefont{F.~M.}
  \bibnamefont{Spedalieri}}, \bibinfo{journal}{Phys. Rev. Lett.}
  \textbf{\bibinfo{volume}{88}}, \bibinfo{pages}{187904}
  (\bibinfo{year}{2002}),
  \urlprefix\url{https://link.aps.org/doi/10.1103/PhysRevLett.88.187904}.

\bibitem[{\citenamefont{Doherty et~al.}(2004)\citenamefont{Doherty, Parrilo,
  and Spedalieri}}]{PhysRevA.69.022308}
\bibinfo{author}{\bibfnamefont{A.~C.} \bibnamefont{Doherty}},
  \bibinfo{author}{\bibfnamefont{P.~A.} \bibnamefont{Parrilo}},
  \bibnamefont{and} \bibinfo{author}{\bibfnamefont{F.~M.}
  \bibnamefont{Spedalieri}}, \bibinfo{journal}{Phys. Rev. A}
  \textbf{\bibinfo{volume}{69}}, \bibinfo{pages}{022308}
  (\bibinfo{year}{2004}),
  \urlprefix\url{https://link.aps.org/doi/10.1103/PhysRevA.69.022308}.

\bibitem[{\citenamefont{Knips et~al.}(2016)\citenamefont{Knips, Schwemmer,
  Klein, Wie\ifmmode~\acute{s}\else \'{s}\fi{}niak, and
  Weinfurter}}]{PhysRevLett.117.210504}
\bibinfo{author}{\bibfnamefont{L.}~\bibnamefont{Knips}},
  \bibinfo{author}{\bibfnamefont{C.}~\bibnamefont{Schwemmer}},
  \bibinfo{author}{\bibfnamefont{N.}~\bibnamefont{Klein}},
  \bibinfo{author}{\bibfnamefont{M.}~\bibnamefont{Wie\ifmmode~\acute{s}\else
  \'{s}\fi{}niak}}, \bibnamefont{and}
  \bibinfo{author}{\bibfnamefont{H.}~\bibnamefont{Weinfurter}},
  \bibinfo{journal}{Phys. Rev. Lett.} \textbf{\bibinfo{volume}{117}},
  \bibinfo{pages}{210504} (\bibinfo{year}{2016}),
  \urlprefix\url{https://link.aps.org/doi/10.1103/PhysRevLett.117.210504}.

\bibitem[{\citenamefont{Shahandeh
  et~al.}(2017{\natexlab{a}})\citenamefont{Shahandeh, Ringbauer, Loredo, and
  Ralph}}]{PhysRevLett.118.110502}
\bibinfo{author}{\bibfnamefont{F.}~\bibnamefont{Shahandeh}},
  \bibinfo{author}{\bibfnamefont{M.}~\bibnamefont{Ringbauer}},
  \bibinfo{author}{\bibfnamefont{J.~C.} \bibnamefont{Loredo}},
  \bibnamefont{and} \bibinfo{author}{\bibfnamefont{T.~C.} \bibnamefont{Ralph}},
  \bibinfo{journal}{Phys. Rev. Lett.} \textbf{\bibinfo{volume}{118}},
  \bibinfo{pages}{110502} (\bibinfo{year}{2017}{\natexlab{a}}),
  \urlprefix\url{https://link.aps.org/doi/10.1103/PhysRevLett.118.110502}.

\bibitem[{\citenamefont{Rosset et~al.}(2012)\citenamefont{Rosset,
  Ferretti-Sch\"obitz, Bancal, Gisin, and Liang}}]{PhysRevA.86.062325}
\bibinfo{author}{\bibfnamefont{D.}~\bibnamefont{Rosset}},
  \bibinfo{author}{\bibfnamefont{R.}~\bibnamefont{Ferretti-Sch\"obitz}},
  \bibinfo{author}{\bibfnamefont{J.-D.} \bibnamefont{Bancal}},
  \bibinfo{author}{\bibfnamefont{N.}~\bibnamefont{Gisin}}, \bibnamefont{and}
  \bibinfo{author}{\bibfnamefont{Y.-C.} \bibnamefont{Liang}},
  \bibinfo{journal}{Phys. Rev. A} \textbf{\bibinfo{volume}{86}},
  \bibinfo{pages}{062325} (\bibinfo{year}{2012}),
  \urlprefix\url{https://link.aps.org/doi/10.1103/PhysRevA.86.062325}.

\bibitem[{\citenamefont{Morelli et~al.}(2022)\citenamefont{Morelli, Yamasaki,
  Huber, and Tavakoli}}]{Morelli2022}
\bibinfo{author}{\bibfnamefont{S.}~\bibnamefont{Morelli}},
  \bibinfo{author}{\bibfnamefont{H.}~\bibnamefont{Yamasaki}},
  \bibinfo{author}{\bibfnamefont{M.}~\bibnamefont{Huber}}, \bibnamefont{and}
  \bibinfo{author}{\bibfnamefont{A.}~\bibnamefont{Tavakoli}},
  \bibinfo{journal}{Phys. Rev. Lett.} \textbf{\bibinfo{volume}{128}},
  \bibinfo{pages}{250501} (\bibinfo{year}{2022}),
  \urlprefix\url{https://link.aps.org/doi/10.1103/PhysRevLett.128.250501}.

\bibitem[{\citenamefont{Buscemi}(2012)}]{PhysRevLett.108.200401}
\bibinfo{author}{\bibfnamefont{F.}~\bibnamefont{Buscemi}},
  \bibinfo{journal}{Phys. Rev. Lett.} \textbf{\bibinfo{volume}{108}},
  \bibinfo{pages}{200401} (\bibinfo{year}{2012}),
  \urlprefix\url{https://link.aps.org/doi/10.1103/PhysRevLett.108.200401}.

\bibitem[{\citenamefont{Branciard et~al.}(2013)\citenamefont{Branciard, Rosset,
  Liang, and Gisin}}]{PhysRevLett.110.060405}
\bibinfo{author}{\bibfnamefont{C.}~\bibnamefont{Branciard}},
  \bibinfo{author}{\bibfnamefont{D.}~\bibnamefont{Rosset}},
  \bibinfo{author}{\bibfnamefont{Y.-C.} \bibnamefont{Liang}}, \bibnamefont{and}
  \bibinfo{author}{\bibfnamefont{N.}~\bibnamefont{Gisin}},
  \bibinfo{journal}{Phys. Rev. Lett.} \textbf{\bibinfo{volume}{110}},
  \bibinfo{pages}{060405} (\bibinfo{year}{2013}),
  \urlprefix\url{https://link.aps.org/doi/10.1103/PhysRevLett.110.060405}.

\bibitem[{\citenamefont{Bancal et~al.}(2011{\natexlab{a}})\citenamefont{Bancal,
  Gisin, Liang, and Pironio}}]{PhysRevLett.106.250404}
\bibinfo{author}{\bibfnamefont{J.-D.} \bibnamefont{Bancal}},
  \bibinfo{author}{\bibfnamefont{N.}~\bibnamefont{Gisin}},
  \bibinfo{author}{\bibfnamefont{Y.-C.} \bibnamefont{Liang}}, \bibnamefont{and}
  \bibinfo{author}{\bibfnamefont{S.}~\bibnamefont{Pironio}},
  \bibinfo{journal}{Phys. Rev. Lett.} \textbf{\bibinfo{volume}{106}},
  \bibinfo{pages}{250404} (\bibinfo{year}{2011}{\natexlab{a}}),
  \urlprefix\url{https://link.aps.org/doi/10.1103/PhysRevLett.106.250404}.

\bibitem[{\citenamefont{Baccari et~al.}(2017)\citenamefont{Baccari, Cavalcanti,
  Wittek, and Ac\'{\i}n}}]{Baccari2017}
\bibinfo{author}{\bibfnamefont{F.}~\bibnamefont{Baccari}},
  \bibinfo{author}{\bibfnamefont{D.}~\bibnamefont{Cavalcanti}},
  \bibinfo{author}{\bibfnamefont{P.}~\bibnamefont{Wittek}}, \bibnamefont{and}
  \bibinfo{author}{\bibfnamefont{A.}~\bibnamefont{Ac\'{\i}n}},
  \bibinfo{journal}{Phys. Rev. X} \textbf{\bibinfo{volume}{7}},
  \bibinfo{pages}{021042} (\bibinfo{year}{2017}),
  \urlprefix\url{https://link.aps.org/doi/10.1103/PhysRevX.7.021042}.

\bibitem[{\citenamefont{Liang et~al.}(2015)\citenamefont{Liang, Rosset, Bancal,
  P\"utz, Barnea, and Gisin}}]{PhysRevLett.114.190401}
\bibinfo{author}{\bibfnamefont{Y.-C.} \bibnamefont{Liang}},
  \bibinfo{author}{\bibfnamefont{D.}~\bibnamefont{Rosset}},
  \bibinfo{author}{\bibfnamefont{J.-D.} \bibnamefont{Bancal}},
  \bibinfo{author}{\bibfnamefont{G.}~\bibnamefont{P\"utz}},
  \bibinfo{author}{\bibfnamefont{T.~J.} \bibnamefont{Barnea}},
  \bibnamefont{and} \bibinfo{author}{\bibfnamefont{N.}~\bibnamefont{Gisin}},
  \bibinfo{journal}{Phys. Rev. Lett.} \textbf{\bibinfo{volume}{114}},
  \bibinfo{pages}{190401} (\bibinfo{year}{2015}),
  \urlprefix\url{https://link.aps.org/doi/10.1103/PhysRevLett.114.190401}.

\bibitem[{\citenamefont{Cavalcanti and
  Terra~Cunha}(2006)}]{doi:10.1063/1.2337535}
\bibinfo{author}{\bibfnamefont{D.}~\bibnamefont{Cavalcanti}} \bibnamefont{and}
  \bibinfo{author}{\bibfnamefont{M.~O.} \bibnamefont{Terra~Cunha}},
  \bibinfo{journal}{Appl. Phys. Lett.} \textbf{\bibinfo{volume}{89}},
  \bibinfo{pages}{084102} (\bibinfo{year}{2006}),
  \eprint{https://doi.org/10.1063/1.2337535},
  \urlprefix\url{https://doi.org/10.1063/1.2337535}.

\bibitem[{\citenamefont{Brand\~ao}(2005)}]{PhysRevA.72.022310}
\bibinfo{author}{\bibfnamefont{F.~G. S.~L.} \bibnamefont{Brand\~ao}},
  \bibinfo{journal}{Phys. Rev. A} \textbf{\bibinfo{volume}{72}},
  \bibinfo{pages}{022310} (\bibinfo{year}{2005}),
  \urlprefix\url{https://link.aps.org/doi/10.1103/PhysRevA.72.022310}.

\bibitem[{\citenamefont{Huber and de~Vicente}(2013)}]{PhysRevLett.110.030501}
\bibinfo{author}{\bibfnamefont{M.}~\bibnamefont{Huber}} \bibnamefont{and}
  \bibinfo{author}{\bibfnamefont{J.~I.} \bibnamefont{de~Vicente}},
  \bibinfo{journal}{Phys. Rev. Lett.} \textbf{\bibinfo{volume}{110}},
  \bibinfo{pages}{030501} (\bibinfo{year}{2013}),
  \urlprefix\url{https://link.aps.org/doi/10.1103/PhysRevLett.110.030501}.

\bibitem[{\citenamefont{Eisert et~al.}(2007)\citenamefont{Eisert, Brand{\~{a}
  }o, and Audenaert}}]{Eisert_2007}
\bibinfo{author}{\bibfnamefont{J.}~\bibnamefont{Eisert}},
  \bibinfo{author}{\bibfnamefont{F.~G. S.~L.} \bibnamefont{Brand{\~{a} }o}},
  \bibnamefont{and} \bibinfo{author}{\bibfnamefont{K.~M.~R.}
  \bibnamefont{Audenaert}}, \bibinfo{journal}{New J. Phys.}
  \textbf{\bibinfo{volume}{9}}, \bibinfo{pages}{46} (\bibinfo{year}{2007}),
  \urlprefix\url{https://doi.org/10.1088%2F1367-2630%2F9%2F3%2F046}.

\bibitem[{\citenamefont{G\"uhne et~al.}(2007)\citenamefont{G\"uhne, Reimpell,
  and Werner}}]{PhysRevLett.98.110502}
\bibinfo{author}{\bibfnamefont{O.}~\bibnamefont{G\"uhne}},
  \bibinfo{author}{\bibfnamefont{M.}~\bibnamefont{Reimpell}}, \bibnamefont{and}
  \bibinfo{author}{\bibfnamefont{R.~F.} \bibnamefont{Werner}},
  \bibinfo{journal}{Phys. Rev. Lett.} \textbf{\bibinfo{volume}{98}},
  \bibinfo{pages}{110502} (\bibinfo{year}{2007}),
  \urlprefix\url{https://link.aps.org/doi/10.1103/PhysRevLett.98.110502}.

\bibitem[{\citenamefont{Shahandeh
  et~al.}(2017{\natexlab{b}})\citenamefont{Shahandeh, Hall, and
  Ralph}}]{PhysRevLett.118.150505}
\bibinfo{author}{\bibfnamefont{F.}~\bibnamefont{Shahandeh}},
  \bibinfo{author}{\bibfnamefont{M.~J.~W.} \bibnamefont{Hall}},
  \bibnamefont{and} \bibinfo{author}{\bibfnamefont{T.~C.} \bibnamefont{Ralph}},
  \bibinfo{journal}{Phys. Rev. Lett.} \textbf{\bibinfo{volume}{118}},
  \bibinfo{pages}{150505} (\bibinfo{year}{2017}{\natexlab{b}}),
  \urlprefix\url{https://link.aps.org/doi/10.1103/PhysRevLett.118.150505}.

\bibitem[{\citenamefont{Rosset et~al.}(2018)\citenamefont{Rosset, Martin,
  Verbanis, Lim, and Thew}}]{PhysRevA.98.052332}
\bibinfo{author}{\bibfnamefont{D.}~\bibnamefont{Rosset}},
  \bibinfo{author}{\bibfnamefont{A.}~\bibnamefont{Martin}},
  \bibinfo{author}{\bibfnamefont{E.}~\bibnamefont{Verbanis}},
  \bibinfo{author}{\bibfnamefont{C.~C.~W.} \bibnamefont{Lim}},
  \bibnamefont{and} \bibinfo{author}{\bibfnamefont{R.}~\bibnamefont{Thew}},
  \bibinfo{journal}{Phys. Rev. A} \textbf{\bibinfo{volume}{98}},
  \bibinfo{pages}{052332} (\bibinfo{year}{2018}),
  \urlprefix\url{https://link.aps.org/doi/10.1103/PhysRevA.98.052332}.

\bibitem[{\citenamefont{Verbanis et~al.}(2016)\citenamefont{Verbanis, Martin,
  Rosset, Lim, Thew, and Zbinden}}]{PhysRevLett.116.190501}
\bibinfo{author}{\bibfnamefont{E.}~\bibnamefont{Verbanis}},
  \bibinfo{author}{\bibfnamefont{A.}~\bibnamefont{Martin}},
  \bibinfo{author}{\bibfnamefont{D.}~\bibnamefont{Rosset}},
  \bibinfo{author}{\bibfnamefont{C.~C.~W.} \bibnamefont{Lim}},
  \bibinfo{author}{\bibfnamefont{R.~T.} \bibnamefont{Thew}}, \bibnamefont{and}
  \bibinfo{author}{\bibfnamefont{H.}~\bibnamefont{Zbinden}},
  \bibinfo{journal}{Phys. Rev. Lett.} \textbf{\bibinfo{volume}{116}},
  \bibinfo{pages}{190501} (\bibinfo{year}{2016}),
  \urlprefix\url{https://link.aps.org/doi/10.1103/PhysRevLett.116.190501}.

\bibitem[{\citenamefont{Moroder et~al.}(2013)\citenamefont{Moroder, Bancal,
  Liang, Hofmann, and G\"uhne}}]{PhysRevLett.111.030501}
\bibinfo{author}{\bibfnamefont{T.}~\bibnamefont{Moroder}},
  \bibinfo{author}{\bibfnamefont{J.-D.} \bibnamefont{Bancal}},
  \bibinfo{author}{\bibfnamefont{Y.-C.} \bibnamefont{Liang}},
  \bibinfo{author}{\bibfnamefont{M.}~\bibnamefont{Hofmann}}, \bibnamefont{and}
  \bibinfo{author}{\bibfnamefont{O.}~\bibnamefont{G\"uhne}},
  \bibinfo{journal}{Phys. Rev. Lett.} \textbf{\bibinfo{volume}{111}},
  \bibinfo{pages}{030501} (\bibinfo{year}{2013}),
  \urlprefix\url{https://link.aps.org/doi/10.1103/PhysRevLett.111.030501}.

\bibitem[{\citenamefont{Chru{\'s}ci{\'n}ski and
  Sarbicki}(2014)}]{chruscinski2014entanglement}
\bibinfo{author}{\bibfnamefont{D.}~\bibnamefont{Chru{\'s}ci{\'n}ski}}
  \bibnamefont{and} \bibinfo{author}{\bibfnamefont{G.}~\bibnamefont{Sarbicki}},
  \bibinfo{journal}{J. Phys. A Math.} \textbf{\bibinfo{volume}{47}},
  \bibinfo{pages}{483001} (\bibinfo{year}{2014}).

\bibitem[{\citenamefont{Vedral et~al.}(1997)\citenamefont{Vedral, Plenio,
  Rippin, and Knight}}]{PhysRevLett.78.2275}
\bibinfo{author}{\bibfnamefont{V.}~\bibnamefont{Vedral}},
  \bibinfo{author}{\bibfnamefont{M.~B.} \bibnamefont{Plenio}},
  \bibinfo{author}{\bibfnamefont{M.~A.} \bibnamefont{Rippin}},
  \bibnamefont{and} \bibinfo{author}{\bibfnamefont{P.~L.}
  \bibnamefont{Knight}}, \bibinfo{journal}{Phys. Rev. Lett.}
  \textbf{\bibinfo{volume}{78}}, \bibinfo{pages}{2275} (\bibinfo{year}{1997}),
  \urlprefix\url{https://link.aps.org/doi/10.1103/PhysRevLett.78.2275}.

\bibitem[{\citenamefont{Bennett et~al.}(1996)\citenamefont{Bennett, DiVincenzo,
  Smolin, and Wootters}}]{PhysRevA.54.3824}
\bibinfo{author}{\bibfnamefont{C.~H.} \bibnamefont{Bennett}},
  \bibinfo{author}{\bibfnamefont{D.~P.} \bibnamefont{DiVincenzo}},
  \bibinfo{author}{\bibfnamefont{J.~A.} \bibnamefont{Smolin}},
  \bibnamefont{and} \bibinfo{author}{\bibfnamefont{W.~K.}
  \bibnamefont{Wootters}}, \bibinfo{journal}{Phys. Rev. A}
  \textbf{\bibinfo{volume}{54}}, \bibinfo{pages}{3824} (\bibinfo{year}{1996}),
  \urlprefix\url{https://link.aps.org/doi/10.1103/PhysRevA.54.3824}.

\bibitem[{\citenamefont{Wootters}(1998)}]{PhysRevLett.80.2245}
\bibinfo{author}{\bibfnamefont{W.~K.} \bibnamefont{Wootters}},
  \bibinfo{journal}{Phys. Rev. Lett.} \textbf{\bibinfo{volume}{80}},
  \bibinfo{pages}{2245} (\bibinfo{year}{1998}),
  \urlprefix\url{https://link.aps.org/doi/10.1103/PhysRevLett.80.2245}.

\bibitem[{\citenamefont{Wei and Goldbart}(2003)}]{PhysRevA.68.042307}
\bibinfo{author}{\bibfnamefont{T.-C.} \bibnamefont{Wei}} \bibnamefont{and}
  \bibinfo{author}{\bibfnamefont{P.~M.} \bibnamefont{Goldbart}},
  \bibinfo{journal}{Phys. Rev. A} \textbf{\bibinfo{volume}{68}},
  \bibinfo{pages}{042307} (\bibinfo{year}{2003}),
  \urlprefix\url{https://link.aps.org/doi/10.1103/PhysRevA.68.042307}.

\bibitem[{\citenamefont{Vidal and Tarrach}(1999)}]{PhysRevA.59.141}
\bibinfo{author}{\bibfnamefont{G.}~\bibnamefont{Vidal}} \bibnamefont{and}
  \bibinfo{author}{\bibfnamefont{R.}~\bibnamefont{Tarrach}},
  \bibinfo{journal}{Phys. Rev. A} \textbf{\bibinfo{volume}{59}},
  \bibinfo{pages}{141} (\bibinfo{year}{1999}),
  \urlprefix\url{https://link.aps.org/doi/10.1103/PhysRevA.59.141}.

\bibitem[{\citenamefont{Terhal and Horodecki}(2000)}]{Terhal2000b}
\bibinfo{author}{\bibfnamefont{B.~M.} \bibnamefont{Terhal}} \bibnamefont{and}
  \bibinfo{author}{\bibfnamefont{P.}~\bibnamefont{Horodecki}},
  \bibinfo{journal}{Phys. Rev. A} \textbf{\bibinfo{volume}{61}},
  \bibinfo{pages}{040301} (\bibinfo{year}{2000}),
  \urlprefix\url{https://link.aps.org/doi/10.1103/PhysRevA.61.040301}.

\bibitem[{\citenamefont{Bourennane et~al.}(2004)\citenamefont{Bourennane, Eibl,
  Kurtsiefer, Gaertner, Weinfurter, G\"uhne, Hyllus, Bru\ss{}, Lewenstein, and
  Sanpera}}]{PhysRevLett.92.087902}
\bibinfo{author}{\bibfnamefont{M.}~\bibnamefont{Bourennane}},
  \bibinfo{author}{\bibfnamefont{M.}~\bibnamefont{Eibl}},
  \bibinfo{author}{\bibfnamefont{C.}~\bibnamefont{Kurtsiefer}},
  \bibinfo{author}{\bibfnamefont{S.}~\bibnamefont{Gaertner}},
  \bibinfo{author}{\bibfnamefont{H.}~\bibnamefont{Weinfurter}},
  \bibinfo{author}{\bibfnamefont{O.}~\bibnamefont{G\"uhne}},
  \bibinfo{author}{\bibfnamefont{P.}~\bibnamefont{Hyllus}},
  \bibinfo{author}{\bibfnamefont{D.}~\bibnamefont{Bru\ss{}}},
  \bibinfo{author}{\bibfnamefont{M.}~\bibnamefont{Lewenstein}},
  \bibnamefont{and} \bibinfo{author}{\bibfnamefont{A.}~\bibnamefont{Sanpera}},
  \bibinfo{journal}{Phys. Rev. Lett.} \textbf{\bibinfo{volume}{92}},
  \bibinfo{pages}{087902} (\bibinfo{year}{2004}),
  \urlprefix\url{https://link.aps.org/doi/10.1103/PhysRevLett.92.087902}.

\bibitem[{\citenamefont{Cerf et~al.}(1999)\citenamefont{Cerf, Adami, and
  Gingrich}}]{PhysRevA.60.898}
\bibinfo{author}{\bibfnamefont{N.~J.} \bibnamefont{Cerf}},
  \bibinfo{author}{\bibfnamefont{C.}~\bibnamefont{Adami}}, \bibnamefont{and}
  \bibinfo{author}{\bibfnamefont{R.~M.} \bibnamefont{Gingrich}},
  \bibinfo{journal}{Phys. Rev. A} \textbf{\bibinfo{volume}{60}},
  \bibinfo{pages}{898} (\bibinfo{year}{1999}),
  \urlprefix\url{https://link.aps.org/doi/10.1103/PhysRevA.60.898}.

\bibitem[{\citenamefont{Horodecki and Horodecki}(1999)}]{PhysRevA.59.4206}
\bibinfo{author}{\bibfnamefont{M.}~\bibnamefont{Horodecki}} \bibnamefont{and}
  \bibinfo{author}{\bibfnamefont{P.}~\bibnamefont{Horodecki}},
  \bibinfo{journal}{Phys. Rev. A} \textbf{\bibinfo{volume}{59}},
  \bibinfo{pages}{4206} (\bibinfo{year}{1999}),
  \urlprefix\url{https://link.aps.org/doi/10.1103/PhysRevA.59.4206}.

\bibitem[{\citenamefont{Lanyon et~al.}(2014)\citenamefont{Lanyon, Zwerger,
  Jurcevic, Hempel, D\"ur, Briegel, Blatt, and Roos}}]{PhysRevLett.112.100403}
\bibinfo{author}{\bibfnamefont{B.~P.} \bibnamefont{Lanyon}},
  \bibinfo{author}{\bibfnamefont{M.}~\bibnamefont{Zwerger}},
  \bibinfo{author}{\bibfnamefont{P.}~\bibnamefont{Jurcevic}},
  \bibinfo{author}{\bibfnamefont{C.}~\bibnamefont{Hempel}},
  \bibinfo{author}{\bibfnamefont{W.}~\bibnamefont{D\"ur}},
  \bibinfo{author}{\bibfnamefont{H.~J.} \bibnamefont{Briegel}},
  \bibinfo{author}{\bibfnamefont{R.}~\bibnamefont{Blatt}}, \bibnamefont{and}
  \bibinfo{author}{\bibfnamefont{C.~F.} \bibnamefont{Roos}},
  \bibinfo{journal}{Phys. Rev. Lett.} \textbf{\bibinfo{volume}{112}},
  \bibinfo{pages}{100403} (\bibinfo{year}{2014}),
  \urlprefix\url{https://link.aps.org/doi/10.1103/PhysRevLett.112.100403}.

\bibitem[{\citenamefont{Navascu\'es et~al.}(2007)\citenamefont{Navascu\'es,
  Pironio, and Ac\'{\i}n}}]{PhysRevLett.98.010401}
\bibinfo{author}{\bibfnamefont{M.}~\bibnamefont{Navascu\'es}},
  \bibinfo{author}{\bibfnamefont{S.}~\bibnamefont{Pironio}}, \bibnamefont{and}
  \bibinfo{author}{\bibfnamefont{A.}~\bibnamefont{Ac\'{\i}n}},
  \bibinfo{journal}{Phys. Rev. Lett.} \textbf{\bibinfo{volume}{98}},
  \bibinfo{pages}{010401} (\bibinfo{year}{2007}),
  \urlprefix\url{https://link.aps.org/doi/10.1103/PhysRevLett.98.010401}.

\bibitem[{\citenamefont{Clauser et~al.}(1969)\citenamefont{Clauser, Horne,
  Shimony, and Holt}}]{PhysRevLett.23.880}
\bibinfo{author}{\bibfnamefont{J.~F.} \bibnamefont{Clauser}},
  \bibinfo{author}{\bibfnamefont{M.~A.} \bibnamefont{Horne}},
  \bibinfo{author}{\bibfnamefont{A.}~\bibnamefont{Shimony}}, \bibnamefont{and}
  \bibinfo{author}{\bibfnamefont{R.~A.} \bibnamefont{Holt}},
  \bibinfo{journal}{Phys. Rev. Lett.} \textbf{\bibinfo{volume}{23}},
  \bibinfo{pages}{880} (\bibinfo{year}{1969}),
  \urlprefix\url{https://link.aps.org/doi/10.1103/PhysRevLett.23.880}.

\bibitem[{\citenamefont{Sperling and Vogel}(2013)}]{PhysRevLett.111.110503}
\bibinfo{author}{\bibfnamefont{J.}~\bibnamefont{Sperling}} \bibnamefont{and}
  \bibinfo{author}{\bibfnamefont{W.}~\bibnamefont{Vogel}},
  \bibinfo{journal}{Phys. Rev. Lett.} \textbf{\bibinfo{volume}{111}},
  \bibinfo{pages}{110503} (\bibinfo{year}{2013}),
  \urlprefix\url{https://link.aps.org/doi/10.1103/PhysRevLett.111.110503}.

\bibitem[{\citenamefont{Gerke et~al.}(2018)\citenamefont{Gerke, Vogel, and
  Sperling}}]{PhysRevX.8.031047}
\bibinfo{author}{\bibfnamefont{S.}~\bibnamefont{Gerke}},
  \bibinfo{author}{\bibfnamefont{W.}~\bibnamefont{Vogel}}, \bibnamefont{and}
  \bibinfo{author}{\bibfnamefont{J.}~\bibnamefont{Sperling}},
  \bibinfo{journal}{Phys. Rev. X} \textbf{\bibinfo{volume}{8}},
  \bibinfo{pages}{031047} (\bibinfo{year}{2018}),
  \urlprefix\url{https://link.aps.org/doi/10.1103/PhysRevX.8.031047}.

\bibitem[{\citenamefont{Svetlichny}(1987)}]{PhysRevD.35.3066}
\bibinfo{author}{\bibfnamefont{G.}~\bibnamefont{Svetlichny}},
  \bibinfo{journal}{Phys. Rev. D} \textbf{\bibinfo{volume}{35}},
  \bibinfo{pages}{3066} (\bibinfo{year}{1987}),
  \urlprefix\url{https://link.aps.org/doi/10.1103/PhysRevD.35.3066}.

\bibitem[{\citenamefont{Collins et~al.}(2002)\citenamefont{Collins, Gisin,
  Popescu, Roberts, and Scarani}}]{PhysRevLett.88.170405}
\bibinfo{author}{\bibfnamefont{D.}~\bibnamefont{Collins}},
  \bibinfo{author}{\bibfnamefont{N.}~\bibnamefont{Gisin}},
  \bibinfo{author}{\bibfnamefont{S.}~\bibnamefont{Popescu}},
  \bibinfo{author}{\bibfnamefont{D.}~\bibnamefont{Roberts}}, \bibnamefont{and}
  \bibinfo{author}{\bibfnamefont{V.}~\bibnamefont{Scarani}},
  \bibinfo{journal}{Phys. Rev. Lett.} \textbf{\bibinfo{volume}{88}},
  \bibinfo{pages}{170405} (\bibinfo{year}{2002}),
  \urlprefix\url{https://link.aps.org/doi/10.1103/PhysRevLett.88.170405}.

\bibitem[{\citenamefont{Bancal et~al.}(2011{\natexlab{b}})\citenamefont{Bancal,
  Brunner, Gisin, and Liang}}]{PhysRevLett.106.020405}
\bibinfo{author}{\bibfnamefont{J.-D.} \bibnamefont{Bancal}},
  \bibinfo{author}{\bibfnamefont{N.}~\bibnamefont{Brunner}},
  \bibinfo{author}{\bibfnamefont{N.}~\bibnamefont{Gisin}}, \bibnamefont{and}
  \bibinfo{author}{\bibfnamefont{Y.-C.} \bibnamefont{Liang}},
  \bibinfo{journal}{Phys. Rev. Lett.} \textbf{\bibinfo{volume}{106}},
  \bibinfo{pages}{020405} (\bibinfo{year}{2011}{\natexlab{b}}),
  \urlprefix\url{https://link.aps.org/doi/10.1103/PhysRevLett.106.020405}.

\bibitem[{\citenamefont{Yu et~al.}(2003)\citenamefont{Yu, Chen, Pan, and
  Zhang}}]{PhysRevLett.90.080401}
\bibinfo{author}{\bibfnamefont{S.}~\bibnamefont{Yu}},
  \bibinfo{author}{\bibfnamefont{Z.-B.} \bibnamefont{Chen}},
  \bibinfo{author}{\bibfnamefont{J.-W.} \bibnamefont{Pan}}, \bibnamefont{and}
  \bibinfo{author}{\bibfnamefont{Y.-D.} \bibnamefont{Zhang}},
  \bibinfo{journal}{Phys. Rev. Lett.} \textbf{\bibinfo{volume}{90}},
  \bibinfo{pages}{080401} (\bibinfo{year}{2003}),
  \urlprefix\url{https://link.aps.org/doi/10.1103/PhysRevLett.90.080401}.

\bibitem[{\citenamefont{Barreiro et~al.}(2013)\citenamefont{Barreiro, Bancal,
  Schindler, Nigg, Hennrich, Monz, Gisin, and Blatt}}]{Ba2013}
\bibinfo{author}{\bibfnamefont{J.~T.} \bibnamefont{Barreiro}},
  \bibinfo{author}{\bibfnamefont{J.-D.} \bibnamefont{Bancal}},
  \bibinfo{author}{\bibfnamefont{P.}~\bibnamefont{Schindler}},
  \bibinfo{author}{\bibfnamefont{D.}~\bibnamefont{Nigg}},
  \bibinfo{author}{\bibfnamefont{M.}~\bibnamefont{Hennrich}},
  \bibinfo{author}{\bibfnamefont{T.}~\bibnamefont{Monz}},
  \bibinfo{author}{\bibfnamefont{N.}~\bibnamefont{Gisin}}, \bibnamefont{and}
  \bibinfo{author}{\bibfnamefont{R.}~\bibnamefont{Blatt}},
  \bibinfo{journal}{Nat. Phys.} \textbf{\bibinfo{volume}{9}},
  \bibinfo{pages}{559} (\bibinfo{year}{2013}),
  \urlprefix\url{https://doi.org/10.1038%2Fnphys2705}.

\bibitem[{\citenamefont{Baccari et~al.}(2019)\citenamefont{Baccari, Tura,
  Fadel, Aloy, Bancal, Sangouard, Lewenstein, Ac\'{\i}n, and
  Augusiak}}]{PhysRevA.100.022121}
\bibinfo{author}{\bibfnamefont{F.}~\bibnamefont{Baccari}},
  \bibinfo{author}{\bibfnamefont{J.}~\bibnamefont{Tura}},
  \bibinfo{author}{\bibfnamefont{M.}~\bibnamefont{Fadel}},
  \bibinfo{author}{\bibfnamefont{A.}~\bibnamefont{Aloy}},
  \bibinfo{author}{\bibfnamefont{J.-D.} \bibnamefont{Bancal}},
  \bibinfo{author}{\bibfnamefont{N.}~\bibnamefont{Sangouard}},
  \bibinfo{author}{\bibfnamefont{M.}~\bibnamefont{Lewenstein}},
  \bibinfo{author}{\bibfnamefont{A.}~\bibnamefont{Ac\'{\i}n}},
  \bibnamefont{and} \bibinfo{author}{\bibfnamefont{R.}~\bibnamefont{Augusiak}},
  \bibinfo{journal}{Phys. Rev. A} \textbf{\bibinfo{volume}{100}},
  \bibinfo{pages}{022121} (\bibinfo{year}{2019}),
  \urlprefix\url{https://link.aps.org/doi/10.1103/PhysRevA.100.022121}.

\bibitem[{\citenamefont{Streltsov}(2009)}]{https://doi.org/10.48550/arxiv.0911.1796}
\bibinfo{author}{\bibfnamefont{A.}~\bibnamefont{Streltsov}},
  \emph{\bibinfo{title}{Geometric measure of entanglement compared to measures
  based on fidelity}} (\bibinfo{year}{2009}),
  \urlprefix\url{https://arxiv.org/abs/0911.1796}.

\bibitem[{\citenamefont{Shapira et~al.}(2006)\citenamefont{Shapira, Shimoni,
  and Biham}}]{PhysRevA.73.044301}
\bibinfo{author}{\bibfnamefont{D.}~\bibnamefont{Shapira}},
  \bibinfo{author}{\bibfnamefont{Y.}~\bibnamefont{Shimoni}}, \bibnamefont{and}
  \bibinfo{author}{\bibfnamefont{O.}~\bibnamefont{Biham}},
  \bibinfo{journal}{Phys. Rev. A} \textbf{\bibinfo{volume}{73}},
  \bibinfo{pages}{044301} (\bibinfo{year}{2006}),
  \urlprefix\url{https://link.aps.org/doi/10.1103/PhysRevA.73.044301}.

\end{thebibliography}

\newpage
\appendix

\newpage

\section{A. Overview of entanglement measures}\label{AppA}

%\comm{Re-write this section as an appendix that explains all the measures}
%Basically, there are  a few approaches to quantify entanglement. One direct approach to define a measure is based on the  minimal distance of concerned state $\rho$ to the set of separable states  denoted by $\Omega$~\cite{PhysRevLett.78.2275},
We discuss three classes of entanglement measures.  Firstly, let $\Omega$ denote the set of all separable states. A generic way quantify the entanglement of a state $\rho$ through its smallest distance to $\Omega$ according to some distance measure $D$ \cite{PhysRevLett.78.2275},
\begin{eqnarray}\label{entmeas}
\textstyle E_{\rm D}=\min_{\varrho\in \Omega}D(\rho, \varrho).
\end{eqnarray}

Some well-known entanglement measures correspond to choosing $D$ as different distance measures.

Choose the relative entropy~\cite{PhysRevLett.78.2275} $S(\rho\|\varrho)=tr(\rho\log \rho)- tr(\rho\log \varrho)$, one has
$$E_{\rm re}\equiv \min_{\varrho\in \Omega} S(\rho\|\varrho), $$ the relative entropy of entanglement acts  an upper bound for the entanglement of distillation.

Arising from Uhlmann's fidelity~\cite{https://doi.org/10.48550/arxiv.0911.1796}  $F(\rho, \varrho)=({\rm Tr}[\sqrt{\sqrt{\rho}\varrho\sqrt{\rho}}])^{2}$, three distance measures are
 the revised geometric measure $E_{\rm if}$, Groverian measure $E_{\rm Gr}$~\cite{PhysRevA.68.042307,PhysRevA.73.044301}, and Bures measure $E_{\rm B}(\rho)$~\cite{PhysRevLett.78.2275} as
\setcounter{equation}{2}
\begin{subequations}
\begin{eqnarray}
&&E_{\rm if}(\rho):=\min_{\varrho \in \Omega} 1-F(\rho, \varrho),\\
&&E_{\rm Gr}(\rho):= \min_{\varrho \in \Omega} \sqrt{1- F(\rho, \varrho)},\\
&&E_{\rm B}(\rho):=\min_{\varrho \in \Omega} 2(1-\sqrt{ F(\rho, \varrho)}).
\end{eqnarray}
\end{subequations}
which have different properties in mathematics. We applies our approach to estimate $E_{\rm if}(\rho)$, and the applications to two other measures are trivial and not shown in text.

% \comm{isnt the second and third red measure trivially the same as the first one? The functions are monotonic in F \armin{the revised geometric measure $E_{\rm if}(\rho):=\min_{\varrho \in \Omega} 1-F(\rho, \varrho)$, Groverian measure~\cite{PhysRevA.68.042307,PhysRevA.73.044301} $E_{\rm Gr}(\rho):= \min_{\varrho \in \Omega} \sqrt{1- F(\rho, \varrho)},$} serves as}

Another natural choice is to consider the trace-distance $D_{\rm tr}(\rho, \varrho):=\frac{1}{2} {\rm Tr}|\rho-\varrho|$  with ${\rm Tr}|A|={\rm Tr}\sqrt{AA^{\dagger}}$, and $$E_{\rm tr}=\min_{\varrho\in \Omega} D_{\rm tr}(\rho,
\varrho).$$

For bipartite states, it is also natural to consider the more refined version of entanglement that corresponds to the Schmidt number of a density matrix. This is the smallest pure-state entanglement dimension needed to prepare $\rho$ via a classical mixture \cite{Terhal2000b}. Substituting $\Omega$ in \eqref{entmeas} with $\Omega_k$, denoting the set of states with Schmidt number $k$, we can equally well view $E_{\rm tr}$ as a quantifier of higher-dimensional entanglement.

Secondly, entanglement measures can be constructed via convex-roofs \cite{PhysRevA.54.3824}.  This amounts to establishing some entanglement measure, $E$, valid for bipartite pure states, and then extend it to mixed states as
\begin{eqnarray}
\tilde E(\rho_{\rm AB})=\min_{\{f_{i}, |\phi_{i}\rangle\}}\sum f_{i} \cdot E(|\phi_{i}\rangle).\label{entcon}
\end{eqnarray}
The minimisation is over all decompositions $\rho_{\rm AB}=\sum_{i}f_{i}\cdot |\phi_{i}\rangle \langle \phi_{i}|$.  Some prominent examples follow.

The entanglement of formation ($\tilde {E}_{\rm F}$), corresponding to choosing $E$ as the von Neumann entropy of the reduced state, namely,  $E_{\rm F}(|\phi\rangle)=-{\rm tr}\phi_{\rm A}\log \phi_{\rm A}$ with $\phi_{\rm A}={\rm Tr}_{\rm B}(|\phi\rangle\langle \phi|)$ ~\cite{PhysRevA.54.3824}.
This measure is  the minimal number of singlets that are needed to build a single copy of the concerned state.

The concurrence ($\tilde {E}_{\rm C}$), for which $E_{\rm C}=\sqrt{2-\rm 2Tr(\phi^{2}_{\rm A})}$ \cite{PhysRevLett.80.2245}, is one of most popular measure due to its explicit expression for a two-qubit system.

The geometric measure of entanglement ($\tilde {E}_{\rm G}$), for which $E_{\rm G}=1-\max_{|\varphi\rangle\in \Omega}|\langle \phi|\varphi\rangle |^{2}$~\cite{PhysRevA.68.042307}.

Thirdly, entanglement can also be quantified by its robustness to the noise~\cite{PhysRevA.59.141, PhysRevA.68.042307}. The  robustness of entanglement
$E_{\operatorname{rob}}(\rho)= \frac{p}{1-p}$
is defined by the  minimal real number $0\leq p\leq 1$ %\comm{unclear referencing. I dont understand what each reference did.}
such that there is a separable   state $\varrho\in \Omega$  such that  $(1-p)\rho+p\varrho$ is separable. This can be generalised by  relaxing the requirement that  $\varrho\in \Omega$ ,  leading  to  the generalized robustness of entanglement denoted by $E_{\rm ROB}$~\cite{PhysRevA.72.022310}.

\section{B. Bounds from $E_{\rm tr}$}\label{AppB}
\begin{enumerate}
\item \emph{ The Proof  of   $E_{\operatorname{re}, \rm F}(\rho)\geq \frac{2w_c^{2}}{\ln 2}$:} By the quantum Pinsker inequality, for arbitrary two states $\rho$ and $\varrho$, one has  $S(\rho\| \varrho)\geq \textstyle \frac{2}{\ln 2} [{D}_{\rm tr}(\rho, \varrho)]^2$. Then we have
 \begin{eqnarray}
E_{\operatorname{re}}&:=&S(\rho||\varrho_{\rm opt})
\geq \textstyle \frac{2}{\ln 2} [D_{\rm tr}(\rho, \varrho_{\rm opt})]^2\nonumber \\
&\geq  &\textstyle \frac{2}{\ln 2} [E_{\operatorname{tr}}(\rho)]^2\geq \frac{2w^2_{c}}{\ln 2}  \nonumber
\end{eqnarray}
where $\varrho_{\rm opt}$ is the closest separable state of $\rho$ with respect to the relative entropy. As $\tilde {E}_{\mathrm{F}}\geq E_{\rm re}$, we immediately have $$\tilde {E}_{\mathrm{F}}\geq  \frac{2w^2_{c}}{\ln 2}.  $$

\item {\em Proof  of  $\tilde {E}_{\rm G}(\rho)\geq w_c^{2}$:} For arbitrary two pure states  $|\phi\rangle$ and $|\phi'\rangle$,  one has  $\sqrt{1-|\langle\phi|\phi'\rangle |^{2}}=\frac{1}{2}\operatorname{tr}\big||\phi\rangle \langle\phi |-|\phi'\rangle \langle\phi' |\big|=D_{\rm tr}(|\phi\rangle, |\phi'\rangle)$. Therefore, for the optimal  decomposition ensemble of $\rho$,  specified by $\{f_{i}, |\phi_{i}\rangle\}$,   achieving the convex roof  of $E_{\rm G}(\rho)$, we have
 \begin{eqnarray}
\textstyle \tilde {E}_{\rm G}(\rho)&=&\sum_{i}f_{i}\cdot E_{\rm G}(\phi_{i})
=\sum_{i}f_{i}[D_{\rm tr}(|\phi_{i}\rangle, |\phi'_{i}\rangle)]^2\nonumber \\
&\geq&[D_{\rm tr}(\rho, \textstyle\sum_{i}f_{i}|\phi'_{i}\rangle\langle \phi'_{i}|)]^2\nonumber
 \geq [E_{{\rm tr}}(\rho)]^2\geq w^{2}_{c},
\end{eqnarray}
where $|\phi'_{i}\rangle$ specifies the closest separable pure state of $|\phi_{i}\rangle$.

\item  {\em Proof of  $\tilde {E}_{\rm C}(\rho)\geq-\sqrt{2}w_c $:} Concurrence for a pure state $|\phi\rangle$  is defined as
$E_{\rm C}(|\phi\rangle)=\sqrt{2-2\rm{tr}\phi_{A}^{2}}=\sqrt{2-2\rm{tr}(|\phi\rangle\langle\phi | \rho_{\phi})}\geq \sqrt{2-2[\operatorname{tr}
(\sqrt{|\phi\rangle\langle\phi|\cdot \rho_{\phi}\cdot|\phi\rangle\langle\phi|})]^{2}} \geq \sqrt{2}D_{\rm tr}(\phi, \rho_{\phi})$ with $\phi_{A}={\rm Tr}_{B}(|\phi\rangle\langle \phi|)$ and $\rho_{\phi}\in \Omega$ the diagonal parts of $|\phi\rangle\langle \phi|$ written in the Schmidt basis,  where we have used   $\operatorname{tr}(\rho_{1}\cdot \rho_{2})\leq [\operatorname{tr}
(\sqrt{\operatorname{\sqrt{\rho_{2}}\rho_{1}\sqrt{\rho_{2}}}})]^{2} $ and the Fuchs-van de Graaf inequality  $\sqrt{1-[\operatorname{tr}
(\sqrt{\operatorname{\sqrt{\rho_{2}}\rho_{1}\sqrt{\rho_{2}}}})]^{2}}\geq D_{\rm tr}(\rho_{1}, \rho_{2})$. Then we have
\begin{eqnarray}
\tilde {E}_{\rm C}(\rho)&\geq & \textstyle \sqrt{2}\min_{\{f_{i}, \phi_{i}\}}D_{\rm tr}(|\phi_{i}\rangle\langle \phi_{i}|, \rho_{\phi_{i}})\nonumber \\
&\geq& \sqrt{2}D_{\rm tr}(\rho, \sum_{i}f_{i}\cdot \rho_{\phi_{i}})\nonumber \\
&\geq& \sqrt{2}\min_{\varrho\in\Omega}D_{\rm tr}(\rho, \varrho)\geq -\sqrt{2}w_{c},\nonumber
\end{eqnarray}
where $\rho_{\phi_{i}}$ specifies the diagonal parts of $|\phi_{i}\rangle\langle \phi_{i}|$ when written in the Schmidt basis.

\item {\em  Proof of $E_{rob, ROB}\geq \frac{-w_{c}}{1+w_{c}}$:} For the $E_{rob, ROB}$, we specify the the optimal state achieving the minimal wight $p$  as $\varrho'$, and $\varrho\equiv (1-p)\rho+p\varrho'\in\Omega$.  Then
    $2p\geq p\cdot \tr|\rho-\varrho'|={\rm Tr}|\rho-\varrho|\geq-2w_{c}$ thus $p\geq -w_{c}$ and we have
     \begin{eqnarray}
E_{\rm rob, ROB}(\rho)\ge\frac{-w_{\rm c}}{1+w_{\rm c}}.
\end{eqnarray}

\item{\em Proof of $E_{\rm Gr}\geq -w_{\rm c}$, $E_{\rm B}\geq2\sqrt{1-\sqrt{1-w^{2}_{\rm c}}}$, and $E_{if}\geq w^{2}_{\rm c}$}.
    All these three measures involve the maximization of infidelity. We only need to prove that $\max_{\varrho \in \Omega} F(\rho, \varrho)\leq 1-w^{2}_{\rm c}$.
    As $\sqrt{1-F(\rho, \varrho)}\geq D_{\rm tr}(\rho, \varrho)$,
    we immediately have
$\min_{\varrho \in \Omega} 1-F(\rho, \varrho)\geq w^{2}_{\rm c}$, and the desired inequalities follows.
 \end{enumerate}

\end{document}